\documentclass[aps,a4paper,reprint,showpacs,amsmath,amssymb,superscriptaddress,floatfix]{revtex4-1}

\usepackage{amssymb}
\usepackage{latexsym}
\usepackage[dvips]{graphicx}
\usepackage{epsfig}
\usepackage{dcolumn}
\usepackage{amsmath}
\usepackage{amsfonts}
\usepackage{bm}
\usepackage{color}

\usepackage{hyperref}
\hypersetup{colorlinks,linkcolor=blue,urlcolor=blue,citecolor=blue}

\newcommand{\be}{\begin{equation}}
\newcommand{\ee}{\end{equation}}
\newcommand{\bea}{\begin{eqnarray}}
\newcommand{\eea}{\end{eqnarray}}

\newcommand{\p}{\partial}
\newcommand{\s}{\sigma}

\newcommand{\la}{\langle}
\newcommand{\ra}{\rangle}
\newcommand{\rd}{\mbox{d}}

\renewcommand{\vec}[1]{{\bm #1}}

\renewcommand\appendixname{Suppl.Mat.}

\begin{document}
\title{Transport in Magnetically Doped One-Dimensional Wires: \\
       Can the Helical Protection Emerge without the Global Helicity?
      }

\author{A. M. Tsvelik}
\affiliation{Condensed Matter Physics and Materials Science Division, Brookhaven National Laboratory, Upton, NY 11973-5000, USA}

\author{O. M. Yevtushenko}
\affiliation{Ludwig Maximilian University, Arnold Sommerfeld Center and Center for Nano-Science,
             Munich, DE-80333, Germany}

\date{\today }

\begin{abstract}
We study the phase diagram and transport properties of arbitrarily doped
quantum wires functionalized by magnetic adatoms. The appropriate theoretical
model for these systems is a dense
one-dimensional Kondo Lattice (KL) which consists of itinerant
electrons interacting with  localized quantum magnetic moments.
We discover the novel phase of the locally helical metal where transport
is protected from a destructive influence of material imperfections.
Paradoxically, such a protection emerges without a need of
the global helicity, which is inherent in all previously studied
helical systems and requires breaking the spin-rotation symmetry.
We explain the physics of this protection of the new type,
find conditions, under which it emerges, and discuss possible experimental
tests. Our results pave the way to the straightforward realization of
the protected ballistic transport in quantum wires made of various materials.
\end{abstract}

\pacs{
   75.30.Hx,   % Magnetic impurity interactions
   71.10.Pm,   % Fermions in reduced dimensions (anyons, composite fermions, Luttinger liquid, etc.)
   72.15.Nj    % Collective modes (e.g., in one-dimensional conductors)
}

\maketitle

\section{Introduction \label{SecIntro}}

Protected states, which are important elements for nanoelectronics, spintronics and
quantum computers, attract evergrowing attention of physicists. A certain protection
strongly reduces effects of material imperfections, including backscattering and localization,
and provides a possibility to sustain the ballistic transport in relatively long samples.

The current progress in understanding protected transport develops in
two directions. The first one is related to time-reversal invariant topological insulators (TIs)
\cite{HasanKane,QiZhang,TI-Shen}. One dimensional (1D) {\it helical} edge modes
of two-dimensional TIs possess  lock-in relation between electron spin and momentum
%%%
% : electrons propagating in opposite directions have opposite spins
%%%
\cite{WuBernevigZhang,XuMoore}.  Though
this locking may protect transport against disorder \cite{Molenkamp-2007,EdgeTransport-Exp0,EdgeTransport-Exp1},
%%%
% Interacting 1D helical electrons are often described by using the Helical Luttinger Liquid (HLL) model
% \cite{Egger_HLL,WuBernevigZhang,Mirlin-HLL,Yevt-Helical}.
%%%
the protection in realistic TIs is not perfectly robust; reasons for this remain an open and intensively debated question
\cite{Molenkamp-2007,EdgeTransport-Exp0,EdgeTransport-Exp1,EdgeTransport-Exp2,AAY_2013,Yevt-Helical,nichele_2016,vayrynen_2016,Klinovaja_Loss_2017,OYeVYu_2019}.

The second direction exploits the emergent helical
protected states in interacting systems which are not necessarily time-reversal invariant.
Numerous examples of suitable interactions include the hyperfine interaction between nuclei moments
and
%%%
% nuclear spins and spins of
%%%
conduction electrons \cite{braunecker_2009b,braunecker_2009a,jk_2013,hsu_2015,aseev_2017},
the spin-orbit interaction in combination with either a magnetic field \cite{streda_2003,pershin_2004} or
with the Coulomb interaction \cite{kainaris_2015,kainaris_2017}, to name just a few; see
Refs.\cite{braunecker_2010,kloeffel_2011,klinovaja_2011a,klinovaja_2011b,klinovaja_2012,pedder_2016}.
State-of-the-art experiments confirm the existence of  helical states governed
by  interactions \cite{quay_2010,scheller_2014,kammhuber_2017,heedt_2017}.

We focus on another recently predicted and very promising 
possibility to realize protected transport in quantum wires functionalized by magnetic adatoms.
The corresponding theoretical model  is a dense 1D Kondo lattice (KL): the 1D array of local
quantum moments -- Kondo impurities (KI) -- interacting with conduction electrons. KLs have been intensively studied
in different contexts, starting from  the Kondo effect and magnetism to the physics of TIs and Tomonaga Luttinger
liquids (TLLs) \cite{tsunetsugu_1997,review-gulacsi,shibata_1999,doniach_1977,read_1984,auerbach_1986,fazekas_1991,sigrist_1992,tsunetsugu_1992,troyer_1993,ueda_1993,Tsvelik_1994,Shibata_1995,ZachEmKiv,shibata_1996,shibata_1997,Honner_1997,sikkema_1997,mcculloch_2002,xavier_2002,white_2002,novais_2002,Novais_2002b,xavier_2003,xavier_2004,yang_2008,smerat_2011,peters_2012,MaciejkoLattice,aynajian_2012,AAY_2013,Yevt-Helical,khait_2018}.
The physics of KL is determined by the competition between the Kondo screening
and the Ruderman-Kittel-Kosuya-Yosida (RKKY) interaction, as illustrated by the famous Doniach's
phase diagram \cite{doniach_1977}.
We have recently predicted that the 1D RKKY-dominated KL with magnetic
anisotropy of the easy-plane type will form a helix spin configuration which gaps out one helical sector of the electrons.
The second helical sector remains gapless. In the resulting helical metal (HM), the disorder induced localization is
parametrically suppressed and, therefore, the ballistic transport acquires a partial protection \cite{TsvYev_2015,Schimmel_2016}.

All previous studies, including the TIs and the interacting helical systems, revealed protection of transport
governed by the global helicity, i.e., helicity of the gapless electrons and/or the spiral spin configuration
were uniquely defined in the
entire sample. The global helicity always requires breaking the spin-rotation symmetry, either internally
(e.g., due to the spin-orbit interaction, or the magnetic anisotropy) or spontaneously (e.g. in relatively
short samples with a strong electrostatic interaction of the electrons). This certainly diminishes experimental
capabilities to fabricate the helical states, especially those governed by the interactions: one always needs
either specially selected materials or a nontrivial fine-tuning of physical parameters. For instance,
the prediction of Refs.\cite{TsvYev_2015,Schimmel_2016} remains practically useless for the experiments
because one can hardly control the magnetic anisotropy.

\begin{figure}[t]
   \includegraphics[width=0.475 \textwidth]{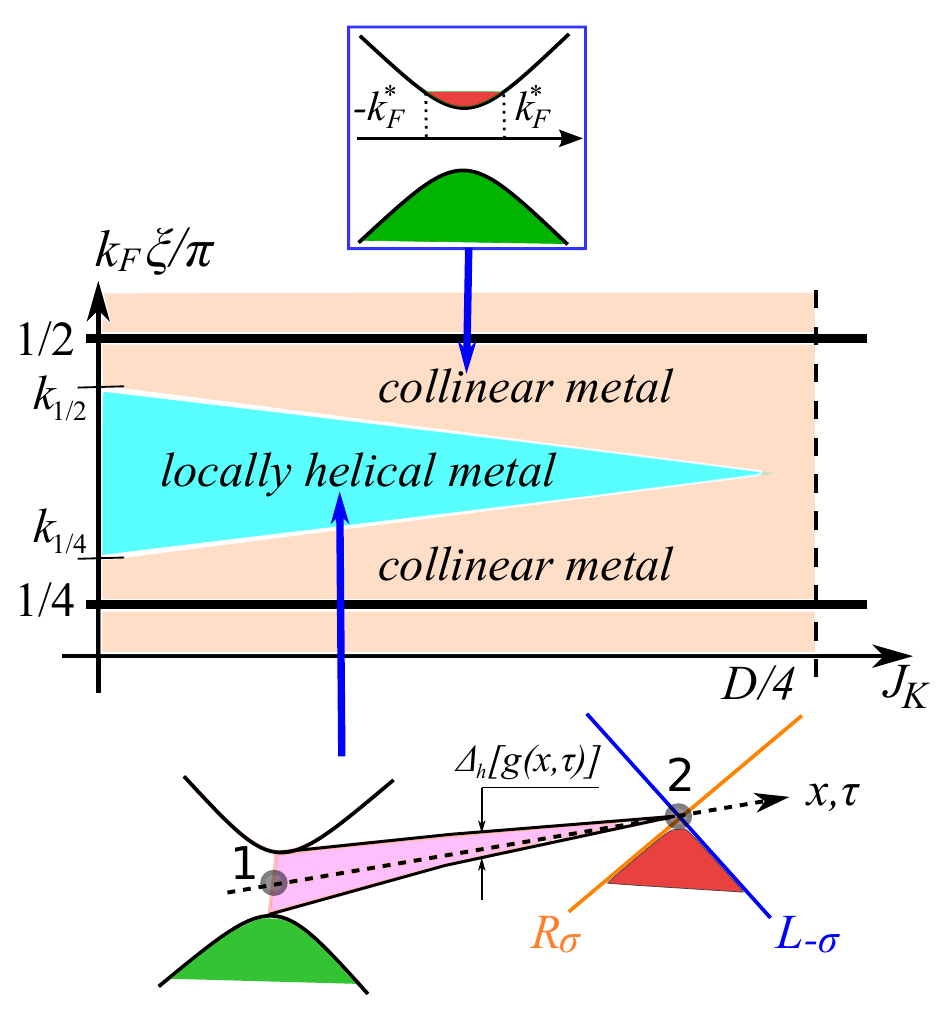}
   \vspace{-0.2cm}
   \caption{
     \label{PhDiagr}
        {\it Central Panel}: Phase diagram of the magnetically doped 1D quantum
        wire for $ J_K \ll E_F $, see
        explanations in the text; here $ | k_f - f \pi / \xi | \sim
        J_K / v_F, \ f = 1/4, 1/2 $.
        {\it Upper panel}: band structure of the
        non-helical collinear metal with the renormalized Fermi-momentum,
        $ k_F^*$. Green and red areas denote filled valence- and partially
        occupied conduction- bands, respectively.
        {\it Lower panel}: band structure and local
        helicity of the novel metallic phase. At some space-time point ``1'',
        the local spin ordering can open a gap, $ \Delta_h $, in the spectrum
        of the fermions $ \{ R_\sigma, L_{-\sigma} \} $ with a given helicity, $ h $. 
        The second helical sector, $ \{ R_{-\sigma}, L_{\sigma} \} $ (not shown
        on this illustration for simplicity), remains gapless at the point ``1''.
        The gap of the fermions $ \{ R, L \} $ slowly varies in space-time due
        to spin fluctuations described by the SU(2) matrix $ g $. There can
        exist another space-time point ``2'' where $ \Delta_h \to 0 $,
        $ | \Delta_{-h} | \to {\rm max} $ and, thus,
        the gapped (gapless) helical sector becomes gapless (gapped). Hence,
        the global helicity cannot be defined though transport remains protected
        as in the case of the globally helical quantum wires.
           }
\vspace{-0.5cm}

\end{figure}

Thus, further progress in obtaining the helical quantum
wires, in particular by means of the magnetic doping, has been  hampered by two open
questions: (i)~Is the global helicity accompanied by breaking
the spin-rotation symmetry really necessary to obtain HM?
(ii)~If the global helicity is not really needed, which parameters must be tuned for
detecting HM in the KLs (theoretically) and in the magnetically doped quantum wires
(experimentally)? We note that
numerical studies have never provided a reliable signature of the helical phase in the
KLs \cite{mcculloch_2002,smerat_2011,khait_2018}.

{\it In this Paper}, we answer both questions:
Protection of the ballistic transport can be provided by the local helicity
which, paradoxically, requires neither the global helicity nor breaking the spin-rotation
symmetry. We show that such a novel HM is the $ 4 k_F $ charge-density-wave (CDW) phase
\cite{Giamarchi} where all effects of disorder are parametrically suppressed. It
can be found in the isotropic KLs if the Kondo exchange coupling is much smaller
than the Fermi energy and the band width, $ J_K \ll E_F, D $,  and the
band filling is far from special commensurate cases (1/4-, 3/4-, 1/2-fillings), see Fig.\ref{PhDiagr}.
To the best of our knowledge, this is the first prediction of the helicity-protected transport in the
quantum 1D system where the spin-rotation symmetry exists and cannot be spontaneously broken.
Our results pave the way towards novel numerical and experimental investigations of the HM.

\section{Theoretical model \label{SectModel}}

We start from the standard KL Hamiltonian:
\be
  \label{model}
  \hat{H} = - \! \sum_n  \Big[
             t \, \psi^\dagger_{n}\psi_{n+1} + h.c. + \mu \, \rho_n  - J_K\psi_n^+ \vec\s \vec{S}_n\psi_n
                                  \Big].
\ee
Here $ \psi_n \equiv \{ \psi_{n,\uparrow}, \psi_{n,\downarrow} \}^{\rm T} $ \ are electron annihilation
($ \psi^\dagger_n $ - creation) operators;  $ \rho_n = \psi^\dagger_{n} \psi_{n} $;
${\vec S}_n$ are quantum  spins with magnitude
$ s $; $ \vec\s \equiv \{ \s_x, \s_y, \s_z \} $ are Pauli matrices;
%%%
% in the spin space;
%%%
$ t $ and $ \mu $ are the electron hopping and the chemical potential;
summation runs over lattice sites.  We assume that $ s J_K < D = 2 t $
and consider low temperatures, $ T \to 0 $.

\section{Method \label{SectMeth}}

We proceed in several steps. Firstly, we find classical
spin configurations minimizing the free energy. Secondly, we identify degrees
of freedom whose fluctuations are gapped, including gapped fermionic and spin
variables ($|{\bf m}|$ and $\alpha$ in Eq.(\ref{SpinDecomp}) below)
and integrate out the gapped variables perturbatively. Remaining spin fluctuations
[described by vectors $\vec{e}_a$ in Eq.(\ref{SpinDecomp})] receive the fully quantum
mechanical treatment. This approach is justified by the separation of scales:
the shortest scale is of order of the inverse Fermi momentum, $1/k_F $. It is
present in the spin ordering and must be much smaller then the coherence length $\zeta$
of the gapped variables. We have performed the self-consistency check which
confirms that $\zeta \gg 1/k_F$ and, thus, justifies the validity of our theory.

\subsection{Separating the slow and the fast variables}

To describe an effective low energy theory, it is convenient to
focus on the regime $ | J_K | < | \mu | \ll t $ where we can linearize the dispersion
relation and introduce right-/left moving fermions, $ \psi_\pm $, in the standard way
\cite{Giamarchi}. In the continuum limit, the fermionic Lagrangian reads
\be
   \label{Lf}
   {\cal L}_F[\psi_\pm] = \sum_{\nu=\pm} \psi^\dagger_\nu \p_\nu \psi_\nu \, ; \quad
   \p_\pm \equiv \p_\tau \mp i v_F \p_x \, .
\ee
Here $ v_F $ is the Fermi velocity, $ \nu $ is the chiral index which indicates the direction
of motion, $ \p_\nu $ is the chiral derivative, $ \tau $ is the imaginary time.

According to Doniach's criterion, the RKKY interaction wins in 1D when the distance
between the spins is smaller then a crossover scale:
$ \, \xi_s < \xi \sqrt{ \vartheta_0 J_K^2/T_K } \, $;  here $ \, \xi \, $ is the lattice spacing, $ \vartheta_0 $
is the density of states at $ E_F $; $T_K$ is the Kondo temperature. We study
this RKKY-dominated regime. For simplicity, we assume
%%%
% will non distinguish lattice constants:
%%%
$ \xi_s = \xi $.

Following Refs.\cite{Tsvelik_1994,TsvYev_2015,Schimmel_2016}, we keep in the Lagrangian
of the electron-KI interaction only the backscattering terms governing the physics of the
dense 1D KL:
%%%
% The corresponding part of the electron-KI interaction on the site $ n $ reads as:
%%%
\be
\label{Lbs}
   {\cal L}_{\rm bs}(n) = J_K \left[ R^\dagger_n \vec\s \vec{S}_n  L_n e^{-2 i k_F x_n} + h.c \right] ,
%%%
% \\
%   R & \equiv & \psi_+,  \, L \equiv \psi_- \, .
%%%
   \ x_n \equiv n \xi \, ;
\ee
where $ R \equiv \psi_+,  \, L \equiv \psi_- $.
$ {\cal L}_{\rm bs} $ contains the fast $ 2 k_F $-oscillations which must be absorbed into the
spin configuration.  We perform this step using the path integral approach where the spin operators
are replaced by integration over a normalized vector field  decomposed as
%%%
% This step can be performed by using a normalized vector field for the
% spin degrees of freedom and applying decomposition
%%%
\bea
   \label{SpinDecomp}
   \vec{S}_n/s = \vec{m} + b \Big(
             & \vec{e}_1 & \, \cos(\alpha) \cos(q x_n + \theta) + \\
             & \vec{e}_2 & \, \sin(\alpha) \sin(q x_n + \theta)
                                          \Bigr)  \sqrt{1 -  \vec{m}^2 } \, .
   \nonumber
\eea
%%%
% Note that $ \vec{m} $ is related to the Kondo physics while $ \vec{e}_{1,2} $
% govern the gap physics.
%%%
Here $ q \simeq 2 k_F $;
$ \{ \vec{e}_1, \vec{e}_2, \vec{m} \} $ is an orthogonal triad of vector fields  whose coordinate
dependence is smooth on the scale $ 1 / k_F $, $ | \vec{e}_{1,2} | = 1 $. Angle $ \alpha $ and
constants $ b,  \theta $ must be chosen to maintain normalization $ | \vec{S}/s | = 1 $. Eq.(\ref{SpinDecomp})
is generic; it allows for only three possible choices of the constants which, in turn,
reflect the band filling $ f $, see \appendixname~A.
After inserting Eq.(\ref{SpinDecomp}) into Eq.(\ref{Lbs}), we select the non-oscillatory  parts of $ {\cal L}_{\rm bs} $
for these three cases, and take the continuous limit. This yields the smooth part of the Lagrangian density:
\begin{itemize}
\item $ f = 1/2, \, 2 k_F x_n = \pi n $:
\be
    \label{half_params}
                                       b = 1, \ \theta = \alpha = 0, \
     {\cal L}_{\rm bs}^{(1/2)} = \tilde{J} \left( \tilde{R}^\dagger \s_x \tilde{L} + h.c \right) ;
\ee

\item $ f = 1/4, \, 2 k_F x_n = \pi n / 2 $:
\bea
    \label{quart_params}
                                       b & = & \sqrt{2}, \ \theta = \pi/4,  \ \alpha \in [0, 2 \pi]; \\
     {\cal L}_{\rm bs}^{(1/4)} & = & \frac{\tilde{J}}{\sqrt{2}} \left( e^{i \pi / 4}
                      \tilde{R}^\dagger [ \cos(\alpha) \s_x + i  \sin(\alpha) \s_y] \tilde{L} + h.c \right) ;
     \nonumber
\eea

\item generic filling:
\bea
  \label{hel_params}
     b = \sqrt{2}, \theta = 0, \alpha = \pi/4 ; \,
     {\cal L}_{\rm bs}^{\rm (gen)} = \tilde{J} \left( \tilde{R}^\dagger \s_{-} \tilde{L} + h.c \right) \!\! .
\eea
\end{itemize}
Here $ \tilde{J} \equiv s J_K \sqrt{1 - \vec{m}^2} / 2 $; $ \s_\pm = (\s_1 \pm i \s_2) / 2 $;  we
expressed vectors $ \vec{e}_{1,2} $ via matrix $ g \in \mbox{SU(2)} $, see \appendixname~B. $ g $
is a smooth function of $ \{ x,  \tau \} $; it governs the new rotated fermionic basis
\bea
 \label{RotatedFerm}
 \tilde{R} & \equiv & g^{-1} R , \ \tilde{L} \equiv g^{-1} L \, ; \\
  {\cal L}_F[\tilde{R},\tilde{L}] & = & \tilde{R}^\dagger ( \p_+ + g^{-1} \p_+ g ) \tilde{R} +
                                                           \tilde{L}^\dagger ( \p_- + g^{-1} \p_- g ) \tilde{L}.
 \nonumber
\eea
Eq.(\ref{half_params}) assumes a staggered configuration of spins at half-filling, $ \uparrow \downarrow $, which
was studied in Ref.\cite{Tsvelik_1994}. The spin sector of the half-filled KL is an antiferromagnet
where the spins fluctuate around the Neel order with a finite correlation length.
Eq.(\ref{quart_params}) reflects two spins up- two spins down configuration,
$ \uparrow \uparrow \downarrow \downarrow $, which agrees with the spin dimerization tendency observed numerically
in Ref.\cite{xavier_2003} at quarter-filling. Eq.(\ref{hel_params}) is a rotationally invariant counterpart of the helical spin
configuration discovered in Refs.\cite{TsvYev_2015,Schimmel_2016} in the anisotropic KL at incommensurate fillings.
Spins fluctuate around this configuration.
Detailed derivation of their effective action is presented in Ref.\cite{KL-SU2-PRB}.
A simplified version of Eq.(\ref{hel_params}) was used in Ref.\cite{fazekas_1991}
for analyzing magnetic properties of KLs.
Below, we refer to Eqs.(\ref{half_params},\ref{quart_params}) at $ \alpha = 0 $ as
``commensurate configurations'' and to Eqs.(\ref{quart_params},\ref{hel_params})  at $ \alpha = \pi/4 $ as
``general configurations''.
We note that the low energy physics of KLs with the 1/4- and 3/4-filling is equivalent in our model. Therefore,
we will discuss only 1/4-filling and do not repeat the same discussion for the case of the 3/4-filling.

\section{Results \label{SectRes}}

Let us start from the presentation of our results at the simplified and transparent
semiclassical level.

\subsection{Fermionic gap}

The backscattering described by Eqs.(\ref{half_params}-\ref{hel_params}) opens a gap
%%%
% at the Dirac point (the level of the chemical potential)
%%%
in the spectrum of the rotated fermions $ \tilde{R}, \tilde{L} $. It
decreases their ground state energy: the larger the gap, the greater is the gain in the fermionic
kinetic energy. Since the spin degrees of freedom do not have kinetic energy,
the minimum of the ground state energy is achieved  by maximizing the fermionic gap. This
indicates that $ | \vec{m} | $ is the gapped variable with the classical value $ m_0 = 0 $.
Below, we use $ m_0 $ for the semiclassical part of the discussion.

The KL contains two fermionic sectors which can have different gaps depending
on the band filling and the spin configuration. The gaps can be found from
Eqs.(\ref{half_params}-\ref{hel_params}):
\bea
   \label{Gap12}
   f=1/2:     & \ &    \Delta^{(1/2)}_{1,2} = \tilde{J} ; \\
   \label{Gap14}
   f = 1/4:     & \ &    \Delta^{(1/4)}_{1,2} = \tilde{J} (\cos(\alpha) \pm \sin(\alpha)) / \sqrt{2} ; \\
   \label{GapHel}
   \mbox{generic filling}: & \ &    \Delta^{\rm (gen)}_{1} = \tilde{J}, \ \Delta^{\rm (gen)}_{2} = 0 .
\eea
The gain in the fermionic ground state energy reads
\be
   \label{DeltaEnGS}
   \delta E_{\rm GS} = - \vartheta_0 \, \xi \sum_{k=1,2} \Delta^2_{k} \log\bigl( D / | \Delta_{k} | ) ;
\ee
see \appendixname~C. $ \vartheta_0 = 1 / \pi v_F $ for the 1D Dirac fermions.  Let
us now analyze various band fillings.

\subsection{Special commensurate fillings, insulating KLs}

At $ f = 1/2, 1/4 $, we have to decide which spin configurations - the commensurate
ones [Eq.(\ref{half_params}) for $ f = 1/2 $ and Eq.(\ref{quart_params}) with $ \alpha = 0, \pi/2 $
for $ f = 1/4 $] or the generic configuration - minimize the ground state energy. Using
Eqs.(\ref{Gap12}-\ref{DeltaEnGS}), we obtain
\bea
  \label{deltaE1/2}
  & &
  \delta E_{\rm GS}(\Delta^{(1/2)}) \!- \! \delta E_{\rm GS}(\Delta^{\rm (gen)}) =
        - {\cal E} \log\left( D / |\tilde{J}| \right) \! , \\
  \label{deltaE1/4}
  & &
   \delta E_{\rm GS}\left( \Delta|^{(1/4)}_{\alpha = 0} \right) \! - \! \delta E_{\rm GS} \left( \Delta|^{(1/4)}_{\alpha = \pi/4} \right) =
       \! - \! {\cal E} \log(\sqrt{2}) ;
\eea
with $ \, {\cal E} \equiv \vartheta_0 \xi \tilde{J}^2 \, $. In both commensurate cases, the commensurate
configuration wins, the conduction band of such KLs is empty and, hence, they are insulators, as expected.
Note that, at quarter-filling, the minimum of $ \delta E_{\rm GS} $ is provided by $ \alpha = 0 $ which
means that $ \alpha $ is gapped.
%%%
% though the gain in energy is less pronounced. Thus, $ \alpha $ is gapped at quarter-filling and its classical
% value correspond to the case where all fermions are gapped and the conduction band is empty -
% KL again is an insulator, as also expected.
%%%

\subsection{Vicinity of special commensurate fillings, collinear metal and heavy TLL}

Let us consider  fillings which are slightly shifted from the special commensurate cases.
To be definite, we analyze an upward shift; a downward shift can be studied in much the same way.
Eqs.(\ref{deltaE1/2}-\ref{deltaE1/4})
suggest that the commensurate spin configuration remain energetically favorable even close to
the commensurate filling.  The wave vector $ q $ of the spin modes remains commensurate,
Eqs.(\ref{half_params},\ref{quart_params}), and is slightly shifted from $ 2 k_F $: $ 2 k_F - q \equiv Q \ll 1/\xi$
with $ q = 2 \pi f / \xi $ and $ f = 1/2, 1/4 $. This case is described in terms of Dirac fermions
with a  non-zero chemical potential:
%%%
% This yields smooth $ Q $-oscillation of the Lagrangians, e.g.:
%%%
\be
     \label{shifted_mu}
%%%
%     \mbox{near 1/2-filling}: \,
%%%
     \bar{{\cal L}} = {\cal L}_F[R,L] + {\cal L}_{\rm bs}^{(f)}[R,L] - (v_F Q / 2 ) (R^\dagger R + L^\dagger L) ,
%%%
%     {\cal L}_{Q}^{(1/2)} \!\!= \tilde{J} \left( e^{-i Q x}\tilde{R}^\dagger \s_x \tilde{L} + h.c \right) ;
%%%
\ee
see \appendixname~D. Backscattering by the commensurate spin configuration opens a gap below the chemical
potential. The electrons with energies $ 0 < E \le v_F Q / 2 $ are pushed above the gap, Fig.\ref{PhDiagr},
and have (almost) parabolic dispersion:
\be
  \label{DispRel}
  E^{+}(k) \bigl|_{v_F |k| < | \tilde{J} | } \simeq |\tilde{J}| + \left( v_F k \right)^2 / 2 |\tilde{J}| \, ;
\ee
see Eq.(D4). Since this new phase possesses a partially filled band it is a metal.
Its metallic behavior originates from the (almost) collinear spin configuration whose classic
component is governed by only one slowly rotating vector, e.g. $ \vec{e}_1 $. We will
reflect this fact by referring to such phases as ``{\it collinear metals}'' (CMs).

A detail description of CMs is presented in Ref.\cite{KL-SU2-PRB}. Let us mention here that
spin modes can mediate repulsion between the conduction electrons
and, for energies  $ | E - E^+(k_F^*) | \ll E^+(k_F^*) $, they  form a repulsive and spinful
TLL characterized by a new Fermi momentum $ k^*_F = Q / 2 $.
%%%
% = k_F - \pi / 2 \xi $ near 1/2-filling or $ k^*_F = Q / 2 = k_F - \pi / 4 \xi $ near 1/4-filling.
%%%
If the effective repulsion is strong enough, TLL becomes {\it heavy}.  Such TLL has been  observed
numerically in Ref.\cite{khait_2018}. 1D nature makes repulsive CMs very sensitive
to spinless impurities: even a weak disorder easily drives it to the localized regime with suppressed
dc transport \cite{GiaSchulz}.

\subsection{Quantum phase transition at generic filling}

The CM becomes less favorable when $ | Q | $ increases: the energy of the electrons in the TLL,
$ E_{\rm p} \simeq \xi \tilde{J} k_F^* / \pi + \xi v_F^2 (k_F^*)^3 / 6  \pi \tilde{J} $, becomes large when
$ k_F^* = Q  / 2 $ increases,  Fig.\ref{PhDiagr}. If $ | Q | $ is large enough, such that $ \, E_{\rm p} \ge {\cal E} \, $,
the minimum of the ground state energy is provided by the generic spin configuration, Eq.(\ref{hel_params}).
Equalizing the leading part of $ E_{\rm p} $ with $ {\cal E} $, we can estimate the critical  value of $ Q $
at which the spin configuration changes: $ Q_c \sim \tilde{J} / v_F $.
If $ \tilde{J} \ll v_F / \xi \sim D $, there is always a parametrically large window of the band fillings where
the new phase is realized,
%%%
% instead of the CM,
%%%
see Fig.\ref{PhDiagr}.
If $ \tilde{J} \to D / 4 $, this window shrinks to zero and the CM dominates at all fillings excluding
special commensurate cases 1/2, 1/4; see Fig.\ref{PhDiagr}.
%%%
% We would like to emphasize that
%%%
The spin configuration cannot change gradually. The switching from the
commensurate to the generic configuration is always abrupt and, therefore, $ Q_c $ is the point of a
quantum phase transition.

\subsection{Generic incommensurate fillings, \\
                   locally helical metal}

The remaining case of generic filling, Eq.(\ref{hel_params}), is the most promising
for transport because rotated fermions are gapped only in one helical sector, e.g. $ \{ \tilde{R}_{\downarrow},
\tilde{L}_{\uparrow} \} $, and the second helical sector, $ \{ \tilde{R}_{\uparrow}, \tilde{L}_{\downarrow} \} $,
remains gapless, see Eq.(\ref{GapHel}) and Fig.\ref{PhDiagr}. The semiclassically broken helical
symmetry is restored by the fluctuations:
The rotating matrix field $g(x,\tau)$ slowly changes in space and time around the underlying spin spiral and,
therefore, the global helicity cannot appear, see Fig.\ref{PhDiagr}.
Hence, one can describe properties of the new phase only in terms
of the local helicity. Simultaneously, there are no sectors of the physical fermions,
$ \{ R_{\sigma}, L_{\sigma} \} $, which can
be found from the inverse of the rotation Eq.(\ref{RotatedFerm}), which are either gapless or globally helical.
To emphasize the underlying {\it locally} helical spin configuration, we refer
to this phase as ``{\it locally-helical metal}'' (lHM).

Since neither the spins nor the physical charge carriers in the lHM possess the global helicity, one
can surmise that they are not a platform for a protected transport. This is, however, incorrect since
the most significant property of the lHMs is that  they
{\it inherit protection of the ballistic transport} from those HMs where SU(2)
symmetry is broken and the global helicity emerges \cite{TsvYev_2015,Schimmel_2016}.
The absence of the global helicity in lMHs is reflected in the gapped nature of the spin excitations \cite{KL-SU2-PRB}

\subsection{Origin of protection}

Let us explain the physics of the seemingly counterintuitive protected transport in the lHMs.
%%%
% We note that
%%%
The density and backscattering operators are invariant under $ g $-rotation:
$ R^\dagger R = \tilde{R}^\dagger \tilde{R}, \, L^\dagger L = \tilde{L}^\dagger \tilde{L}, \,
R^\dagger L = \tilde{R}^\dagger \tilde{L} $. The low energy physics is governed by fields whose correlation functions
decay as  power law. To obtain them, we project the fields on the gapless sector, i.e., average over the high energy gapped
modes. For example, components of the charge density are:
\bea
             \rho (0) & = & \tilde{R}^\dagger_{\uparrow} \tilde{R}_{\uparrow} + \tilde{L}^\dagger_{\downarrow} \tilde{L}_{\downarrow} \, ; \cr
             \rho (4 k_F) & \sim &
       e^{- 4 i k_F x} \tilde{R}^\dagger_{\uparrow} \langle \tilde{R}^\dagger_{\downarrow} \tilde{L}_{\uparrow} \rangle \tilde{L}_{\downarrow}
      + h.c.
\nonumber      
\eea
$ \rho (2 k_F) $ is absent because it would correspond to a single particle
elastic backscattering between the gapless and gapped fermions which is not allowed.
This is the direct consequence of the (local) spin helix which gaps out only one helical fermionic sector.
%%%
% The same arguments are used to obtain the expression for $ \rho(4 k_F) $.
%%%
Thus, the HM is the $ 4 k_F $-CDW phase. This fact has two important consequences: (i) the (local) spin
helix shifts the Friedel oscillations of the charge density from $ \, 2k_F \, $ to $ \, 4k_F \, $, which is indistinguishable
from $ \, 4 (k_F - \pi/2 \xi ) \, $ due to the lattice periodicity; (ii) even more importantly, it drastically reduces
backscattering caused by spinless disorder.
%%%
% {\color{blue} The later effect comes from the fact that only $4k_F$ response is singular, but the amplitude the corresponding
% CDW is proportional to $ \langle \tilde{R}^\dagger_{\downarrow} \tilde{L}_{\uparrow} \rangle  \sim \Delta/E_F$ and hence is
% drastically reduced.
% }
%%%

To illustrate the 2nd statement, we introduce a random potential of spinless backscattering impurities which
couples to the $ 2 k_F$-component of density $  {\cal L}_{\rm dis} = V_{2 k_F} \, \tilde{R}^\dagger \tilde{L} + h.c. $
Here $ V_{2 k_F} $ is a smooth $ 2 k_F$-component of the random potential. Since the charge response function
of lHM at the $ 2k_F$ wave-vector is non-singular, backscattering can occur only via many particle processes with
much smaller amplitude. Averaging over the gapped fermions, we find:
$
   \langle {\cal L}_{\rm dis} \rangle \simeq 2  \left( V_{2 k_F}^2 / \Delta^{\rm (gen)} \right) \,
                                                                                    \tilde{R}^\dagger_{\uparrow} \tilde{L}_{\downarrow} + h.c.
$
see \appendixname~E. If the helical gap is large enough, $ \Delta^{\rm (gen)} =
\tilde{J} \gg V_{2 k_F} $, backscattering and all disorder effects are parametrically suppressed.

%%%
% {\it A brief summary of full quantum-mechanical theory}:
% For completeness, we summarize the key steps of the full quantum-mechanical theory,
% see details in Ref.\cite{KL-SU2-PRB}.
%
% Firstly, we integrate out gapped fermions, exponentiate the fermionic
% determinant and expand the full Lagrangian in gradients of the matrix $ g $ and in small
% fluctuations of $ | m | $ around its classical value. Next, we reinstate the Wess-Zumino
% term for the spin field \cite{ATsBook} and select its smooth parts. The commensurate spin
% configurations generate also the topological term (see Ref.\cite{Tsvelik_1994}, Sect.16 of the book
% \cite{ATsBook}, and references therein). Finally, we integrate out fluctuations of $ | m | $
% in the Gaussian approximation. These steps result in nonlinear $ \sigma $-models
% in 1+1 dimensions, which describe the spin degrees of freedom and suggest stability
% of the mean-field predictions.
%%%

\section{Quantum theory for smooth spin variables and self-consistency check
            \label{SectSigmaModel}}

To complete the theory of the magnetically doped quantum wires, one must consider
quantum fluctuations of smooth spin variables $ \vec{e}_{1,2} $. They
are described by using the heavy field-theoretical machinery of the nonlinear
$ \sigma $-model (nLSM). Its derivation is a lengthy task
which is described in detail in Ref.\cite{KL-SU2-PRB}. Here, we very briefly recapitulate
main steps of the derivation, give final answers, and argue that the fully quantum mechanical
theory does not violate separation of scales, see Sect.\ref{SectMeth}. The
latter is especially important since it confirms validity of our approach and validates
results described in the previous Section at the simplified and transparent semiclassical level.

Derivation of the nLSM requires several steps:

\begin{itemize}

\item
One (i) integrates out gapped fermions and exponentiates the fermionic determinant; (ii)
derives the Jacobian of the SU(2) rotation by the matrix $ g $; (iii) selects smooth contributions
from the Wess-Zumino term for the spin field \cite{ATsBook}. The commensurate spin configurations
generate also the topological term (see Ref.\cite{Tsvelik_1994}, Sect.16 of the book \cite{ATsBook},
and references therein).

\item
The total Lagrangian, which is obtained by summing up the exponentiated fermionic
determinant, the Jacobian, the Wess-Zumino contributions and the topological terms,
is expanded in gradients of the matrix $ g $ and in small fluctuations of $ | \vec{m} | $ around
its classical value $ m_0 = 0 $. The commensurate spin configuration, which corresponds
to 1/4-filling, requires also the expansion in fluctuations of $ \alpha $.

\item
Finally, fluctuations of $ | \vec{m} | $ (and of $ \alpha $, if needed) are integrated out in the
Gaussian approximation.

\end{itemize}

These steps result in the quantum mechanical nLSM in (1+1) space-time
dimensions which describes the smooth spin degrees of freedom. Our approach is
self-consistent if typical scales of the quantum theory remain large, $ \ge v_F / \tilde{J} \gg k_F^{-1} $.
The nLSM is different in different phases.

{\it Commensurate insulators and collinear metals}: The action of the nLSM describing fluctuations
of the spin variables in a commensurate insulator and in a collinear metal takes the following form:
\bea
   S^{(f)} & = & \int {\rm d} \tau {\rm d} x {\cal L}^{(f)} + S_{\rm top} , \, S_{\rm top} = (2 s - 1 ) i \pi k \, ; \\
  {\cal L}^{(f)} & = & \frac{1}{2 g_{f}}
                  \left[
                     \frac{(\p_\tau \vec{e}_1)^2}{c_{f}}  +  c_{f} (\p_x \vec{e}_1)^2
                  \right] .
\nonumber
\eea
Here $ f = 1/2 $ at (or close to) the half-filling and $ f = 1/4 $ at (or close to) the quarter-filling;
small dimensionless coupling constants, $ g_{1/2} \simeq (4 \pi / s ) \, \vartheta_0 \xi \tilde{J}
\sqrt{ \log\Bigl( D/|\tilde{J}| \Bigr) } \ll 1 $ and $ g_{1/4} \simeq g_{1/2} / \sqrt{8} \ll 1 $,
determine small renormalized velocities of the spin excitations, $ c_{f} = v_F g_{f} / 4 \pi \ll v_F $.
Smallness of $ g_f $ and $ c_f $ reflects the coupling between spins and gapped (localized) fermions.
The integer $ k $  marks topologically different sectors of the theory.

The action $ S^{(f)} $ corresponds to the well-known O(3)-symmetric nLSM in (1+1) dimensions with the
topological term. It is exactly solvable \cite{Wiegmann_1985,Fateev_1991,ATsBook} and possesses
a characteristic energy $ {\cal E}_{f} \sim | \tilde{J} | g_f^{-1}\exp(- 2\pi/g_f) $ which governs a large
spatial scale: $ c_f / {\cal E}_{f} \gg v_F / \tilde{J} \gg k_F^{-1} $. The latter inequality confirms validity
of our approach.

{\it Locally-Helical metals}: The Largangian of the nLSM describing fluctuations
of the spin variables in a lHM takes the following form:
\be
  {\cal L}^{\rm (hel)} = \frac{1}{2 g_{\rm hel}}
                  \left\{
                     \frac{ \left[ \Omega_\tau^{(z)} \right]^2 }{c_{\rm hel}}  +  c_{\rm hel} {\rm tr}( \p_x g^+ \p_x g)
                  \right\} \, ;
\ee
with $ g_{\rm gen} \simeq g_{1/4} / 4 \ll 1 $, $ c_{\rm gen} = v_F g_{\rm gen} / \pi \ll v_F $, and
$ \Omega_\tau^{(z)} \equiv i {\rm tr}[ \s_{z} g^{-1} \p_\tau g] / 2 $. This theory is anisotropic and
has the SU(2)-symmetry, $ g \to {\cal M} g, \, {\cal M} \in \mbox{ SU(2)} $. The time
derivative is present only in the $\Omega^z$ term. This points to a relatively short bare correlation
length of spins which coincides with the UV cut-off of the theory. The latter is $ \sim v_F / \tilde{J} $
in our approach and does not violate the self-consistency requirement because $ v_F / \tilde{J} \gg
\xi $. The actual shortest scale of the theory is expected to be much larger if the anisotropy is irrelevant
and $ {\cal L}^{\rm (gen)} $ flows in the IR limit to the well-known SU(2)$ \times $SU(2)-symmetric nLSM.
An example of such a behaviour is provided by the RG equations derived in Ref.\cite{azaria_1992}
for the (2+1) dimensions. There is no counterargument against the irrelevance of the anisotropy
in the (1+1) dimensions. Therefore, we arrive at a conclusion that the shortest spatial scale generated
by $ {\cal L}^{\rm (hel)} $ is $ \gg v_F / \tilde{J} \gg k_F^{-1} $.

This concludes the self-consistency check of our approach and justifies qualitative results
described in Sect.\ref{SectRes} at the semiclassical level.

\section{Possible numerical and experimental test  of our theory \label{SectTests}}

An important task for the subsequent research is to reliably detect different metallic phases
in the 1D KLs (numerically) and in the magnetically doped quantum wires (experimentally).
This requires to tune  the band filling and the Kondo
coupling. The key features distinguishing CM and lHM in
numerics and experiments are as follows. The conductance of the CM is equal to the quantum $ G_0 = 2 e^2 / h $
while the lHM must show only $ G_0 /2 $ conductance due to the helical gap. %lifted spin degeneracy.
The CM is a spinful TLL whose charge and spin response functions have a peak at $ 2k_F^* $;
$ k_F^* $ is the shifted Fermi momentum predicted by general theorems \cite{yamanaka_1997,oshikawa_2000}.
The lHM is the $4k_F$-CDW and has singular response in the charge sector. Since $4k_F$ and
$ 4k_F^*$ are indistinguishable on the lattice the response of the lHM does
not show the shift $ k_F \to k_F^* $. Unlike systems with broken SU(2) 
symmetry \cite{TsvYev_2015,Schimmel_2016}, the lHM, which we have considered, does not have singular
response in the spin sector. Inasmuch as the CM responds  to scalar potentials at $2k_F^*$ and the lHM -
at $ 4k_F $, the spinless disorder potential has a profound difference with respect to
transport in the CM and lHM phases. Namely, localization is parametrically suppressed in the lHM.

Detecting the CM is not difficult because it is generic at relatively
large $ J_K $ and filling away from 1/2, 1/4. The heavy TLL, which is formed by
the interactions in the CM,  has been observed in numerical
results of Ref.\cite{khait_2018}. However, $ J_K $ was too large for finding the HM. The KL
studied in Ref.\cite{smerat_2011} exhibits an unexpected $ 2 k_F $-peak at small $ J_K $.
Yet, the peak was detected in the spin susceptibility of 6 fermions
distributed over 48 sites. So small KL cannot yield a conclusive support or disproof
of our theory. A more comprehensive study of the larger KLs is definitely needed.

The thorough control of the system parameters is provided by the experimental laboratory of cold
atoms where 1D KL was recently realized \cite{Riegger_2018}. Experiments in cold atoms are, probably,
the best opportunity to test our theory. However, modern solid-state technology also allows one
to engineer specific 1D KL even in solid state platforms. It looks feasible to fabricate 1D KL
in clean 1D quantum wires made, e.g., in GaAs/AlGaAs by using cleaved edge overgrowth technique
\cite{CEO} or in SiGe \cite{mizokuchi_2018}. Magnetic adatoms can be deposited close
to the quantum wire by using the precise ion beam irradiation. One can tune parameters of
these artificial KLs by changing the gate voltage, type and density of the magnetic adatoms and
their proximity to the quantum wire.
Such a nano-engineering of 1D KL is essentially similar to the successful realization of topological
superconductivity in atomic chains \cite{feldman_2016}, in carbon nanotubes \cite{desjardins_2019},
and in Bi \cite{jack_2019}. The experiments should be conducted at low
temperatures, $ T \ll \Delta, {\cal E} $, where destructive thermal fluctuations are weak.

\section{Conclusions \label{SectConcl}}

We have studied the physics of quantum wires functionalized
by magnetic adatoms with a high density and a small coupling between the itinerant electrons
and local magnetic moments of the ad-atoms, $ |J_K| \ll E_F $. Their physics is determined by the
RKKY interaction between the ad-atoms which results in a quite rich phase diagram. It includes:
(i) the insulating phase which appears at special commensurate band filling, either 1/2, or 1/4,
3/4; (ii) spinful interacting metals which exist in the vicinity of that commensurate fillings;
and (iii) the novel metallic phase at generic band fillings, see Fig.\ref{PhDiagr}.

The third phase is our most important and intriguing finding. On one hand, the local spins form
a slow varying in space and time spiral, which can yield a local helical gap of the electrons.
On the other hand, the global helicity is absent because the spin-rotation symmetry is not (and
cannot) be broken. The latter can result in an erroneous conclusion that a helicity-protected
transport could not originate in these locally helical metals. That is not true:
paradoxically, the locally helical phase inherits protection of the ballistic transport from
those systems where the spin rotation symmetry is broken and the global helicity emerges.
Protection of transport in lHMs has a simple physical explanation because they are the $4k_F$-CDW
phase with the reduced $ 2 k_F $ response. This reduction is the direct consequence
of the local helicity. It parametrically suppresses effects of a spinless disorder and localization.
Thus, we come across the principally new type of
emergent (partial) protection of transport caused by the interactions
without a need of the global helicity.
Our model and approach allow us to uncover the promising possibility for engineering the HM in
the quantum wires and to identify the parameter range where the HM is formed, see Fig.\ref{PhDiagr}.
To the best of our knowledge, this gives the firstever example of such a protection in the system
where the spin-rotation symmetry is not (and cannot be) broken. It would be interesting to study
in the future how the direct Heisenberg interaction between the spins could modify out theory \cite{KH-CSL,TI-SupSol}.

We believe that detecting the lHMs in numerical simulations and
real experiments is the task of a high importance. Our results suggest how to tune
the physical parameters, in particular the band filling and the Kondo coupling, such
that the lHM could be realized. The fundamental sensitivity of the state and of the
transport properties of the magnetically doped quantum wire to the band filling is
especially important. It allows one to switch over normal and locally helical
regimes of the conductor by varying a gate voltage. This can be used for creating fully
controllable helical elements. Our theoretical prediction, that the backscattering is
suppressed in the lHMs in spite of the absence of the global helicity,
can pave the way towards flexible engineering principally
new units of nano-electronics and spintronics
%%%
% which could operate as elements of green technology
%%%
with substantially improved efficiency.

%%%
% \hspace{0.25 cm}
%
% {\it Author contributions}: The authors have made an equal contribution to this paper.
%%%

\begin{acknowledgments}
We are grateful to Jelena Klinovaja for useful discussions.
A.M.T. was supported by the U.S. Department of Energy (DOE), Division of Materials Science,
under Contract No. DE-SC0012704. O.M.Ye. acknowledges support from the DFG through the
grants YE 157/2-1\&2. We acknowledge hospitality of the Abdus Salam ICTP
where the part of this project was done. A.M.T. also acknowledges the hospitality of
%%%
% Department of Physics of
Ludwig Maximilian University Munich where this paper was finalized.
%%%
% , and the Cluster of Excellence, Nanosystems Initiative Munich.
%%%
\end{acknowledgments}

%%%
% \newpage
%%%

\bibliography{Bibliography,KL,HeliPhys}

%%%

\widetext

\newpage

\begin{center}
  {\large
    {\bf
     Supplemental Materials for the paper \\ \vspace{0.25cm}
    }
     ``Transport in Magnetically Doped One-Dimensional Wires''
  } \\ \vspace{0.25cm}
     by A. M. Tsvelik and O. M. Yevtushenko
\end{center}

\appendix

\section{Decomposition of a normalized vector field into constant and
              oscillating parts \label{DecompApp}}

Let us consider a unit-vector field, $ \vec{s} $ with
$ | \vec{s} | = 1 $, and single out its zero mode and $ \pm q $ components:
\be
  \label{FieldDecomp}
  \vec{s} = \vec{s}_0 + \vec{s}_c \cos(q x + \theta) + \vec{s}_s \sin(q x + \theta) \, .
\ee
Here $ \theta $ is a constant phase shift; coefficients $ \vec{s}_{0,c,s} $ must be
smooth functions on the scale of $ 1 / q $.
%%%
% the lattice spacing.
%%%
The normalization of $ \vec{s} $ must hold true for arbitrary $ x $. This
{\it always} requires mutual orthogonality
\be
  (\vec{s}_0,\vec{s}_c) = (\vec{s}_0,\vec{s}_s) = (\vec{s}_c,\vec{s}_s) = 0 \, ;
\ee
and proper normalizations:
\bea
   \label{HelConf}
   \mbox{generic $ q $}: & \quad & | \vec{s}_c | = | \vec{s}_s | , \
                                        | \vec{s}_0 |^2 + | \vec{s}_c |^2 = 1 \, ; \\
   \label{OneHalfConf}
   \sin(q x+ \theta)=0:      & \quad & | \vec{s}_0 |^2 + | \vec{s}_c |^2 = 1 \, , \qquad
   \mbox{ or } \quad
   \cos(q x+ \theta)=0:    \qquad | \vec{s}_0 |^2 + | \vec{s}_s |^2 = 1 \, ; \\
   \label{OneQuartConf}
   e^{ i (q x+ \theta )} = \pm \frac{1 \pm i}{\sqrt{2}}:
                                       & \quad &
            | \vec{s}_0 |^2 + \frac{ |\vec{s}_c|^2 + |\vec{s}_s|^2}{2} = 1 \, .
\eea
There are no other configurations which are compatible with decomposition
Eq.(\ref{FieldDecomp}).

\section{Useful relations \label{UslRel}}

Using the matrix identities
\be
\label{MatrId}
 \left\{
 \begin{array}{l}
  \hat{A} = A^{(j)} \sigma_j, \quad  A^{(j)} = \frac{1}{2} {\rm tr}[\sigma_j \hat{A}]; \\ \\
  {\rm tr}[ \vec{\sigma} \hat{A}^{-1} \sigma_j \hat{A}] \, {\rm tr}[ \vec{\sigma} \hat{A}^{-1} \sigma_{j'} \hat{A}] =
        4 \delta_{j,j'}
 \end{array}
 \right. \qquad
 j, j' = x, y, z.
\ee
and re-parameterizing the (real) orthogonal basis $ \vec{e}_{1,2,3} $ in terms of a matrix $ g \in \mbox{SU(2)} $:
\be
\label{BasisFromSU2}
   \vec{e}_{1,2,3} = \frac{1}{2} {\rm tr}[ \vec{\sigma} g \sigma_{x,y,z} g^{-1}]  \, , \quad
   \vec{e}_3 = [ \vec{e}_1 \times  \vec{e}_2 ] \, , \quad
    \sum_{a=1,2,3} ( \p_\alpha \vec{e}_a)^2 = 4 {\rm tr} [ \p_\alpha g^{-1} \p_\alpha g ] \, ;
\ee
we can re-write a scalar product $ ( \vec{\s}, e_j ) $ as follows:
\be
    \label{Vec_SU2}
    ( \vec{\s}, \vec{e}_{1,2} )                                 = \frac{1}{2} g \sigma_{x,y} g^{-1}
       \quad \Rightarrow \quad
    ( \vec{\s}, [ \vec{e}_{1} \pm  i \vec{e}_{1} ] )  = g \sigma_{\pm} g^{-1} \, ; \quad
     \sigma_{\pm} \equiv ( \s_x \pm i \s_y ) / 2.
\ee
One can also do an inverse step and express the SU(2) matrix via a unit vector
\be
  g = i (\vec{\s},\vec{n}), \ g^{-1} = - i (\vec{\s},\vec{n}) \,; \ | \vec{n} | = 1 \, \quad \Rightarrow \quad
  g^{-1} \p_\alpha g = i \bigr( \vec{\s}, [ \vec{n} \times \p_\alpha \vec{n} ] \bigl) \, .
\ee

\section{Ground state energy of the gapped 1D Dirac fermions \label{GSenergy}}

Consider 1D Dirac fermions with the inverse Green's function:
\be
   [\hat{G}(\Delta)]^{-1}
        = \left(
             \begin{array}{cc}
                \p_+    & \Delta \\
                \Delta & \p_-
             \end{array}
           \right)  \, \underrightarrow{\rm FT} \,
          \left(
             \begin{array}{cc}
                -i \omega_n + v_F k    & \Delta \\
                \Delta                           &  - i \omega_n - v_F k
             \end{array}
           \right) .
%%%
%          \quad \p_\pm \equiv \p_\tau \mp i v_F \p_x \, .
%%%
\ee
Integrating out the fermions we find the partition function:
\be
   Z[\Delta] = Z_0 \frac{\det \left( [\hat{G}(\Delta)]^{-1} \right)} {\det \left( [\hat{G}_0]^{-1} \right) } =
                      Z_0 \exp\left( - {\rm Tr} \left( \log \left[  \hat{G}_0^{-1} \hat{G}(\Delta) \right] \right) \right) \simeq
                      Z_0 \exp\left( - {\rm Tr} \left[  \hat{G}_0^{-1} \hat{G}(\Delta)  - 1 \right] \right)  \, .
\ee
Here $ Z_0 \equiv Z[\Delta=0] $, $ \hat{G}_0 \equiv \hat{G}[\Delta=0] $ and $ \Delta $ is assumed
to be small. Using the expression for the free energy $ {\cal F} = - T \log[Z] $, we find that the
gain of the energy, which is caused by the gap opening, reads as
\be
   \delta E _{\rm GS} = T \, {\rm Tr} \left[  \hat{G}_0^{-1} \hat{G}(\Delta)  - 1 \right]
\ee
At $ T = 0 $ and in the continuous limit, this expression reduces to
\be
   \delta E_{\rm GS} = - 2 \xi \int \frac{{\rm d}^2 \{ \omega, q \}}{(2 \pi)^2}
                                                  \frac{\Delta^2}{\omega^2 + (v_F q)^2 + \Delta^2} \, .
\ee
The UV divergence must be cut by the band width $ D $. Thus, we obtain with the logarithmic accuracy:
\be
   \delta E_{\rm GS} \simeq - \frac{\xi}{\pi v_F } \Delta^2 \log\bigl( D / | \Delta | ) \, .
\ee

\section{Smoothly oscillating backscattering \label{ShiftedGap}}

The theory close to the special commensurate filling can be formulated in terms of Dirac fermions
with a spatially oscillating backscattering described by Lagrangian:
\be
 \label{L_osc}
   {\cal L}_{\rm osc} =
   \left(
     \begin{array}{cc}
                    \p_+                & J e^{- i Q x} \\
                    J e^{i Q x}     & \p_-
     \end{array}
   \right) .
\ee
The wave vector $ Q $ is a deviation of $ 2 k_F $ from its special commensurate value.
By rotating the fermions
\be
   R \to e^{-i Q x/ 2} R, \ L \to e^{i Q x/ 2} L,
\ee
we reduce $ {\cal L}_{\rm osc} $ to the Lagrangian with the constant backscattering and with the shifted
chemical potential:
\be
   \label{Losc_rot}
   \bar{{\cal L}}_{\rm osc} =
   \left(
     \begin{array}{cc}
         - i \omega_n + v_F k       & J  \\
        J                           & - i \omega_n - v_F k
     \end{array}
   \right) - \frac{v_F Q}{2}.
\ee
Backscattering opens the gap in the fermionic spectrum but at the energy level shifted from zero
by $ v_F Q / 2 $. Thus, the dispersion relation counted from the shifted chemical potential reads as
\be
   \label{EnOsc}
  J \ne 0 \ \Rightarrow \ E_{\rm osc}^{\pm}(k) = \pm \sqrt{ \left( v_F k \right)^2 + J^2} \Bigl|_{v_F |q| \ll |J|}
                    \simeq \pm \left( |J| + \frac{\left( v_F k \right)^2}{2 |J|} \right) \, .
\ee

%%%
% Note that Eq.(\ref{EnOsc}) is invalid at $ J = 0 $.
%%%

\section{$ 4 k_F $-response of the helical metal on spinless disorder \label{Dis-4KF}}

Consider a $4k_F$-response of the helical metal on the spinless backscattering potential. It requires a fusion of two
$2k_F$-operators which is obtained in path integral by integrating out the high energy gapped modes. The effective
Lagrangian reads as:
\bea
 \langle {\cal L}_{\rm dis} \rangle & = & - \frac{1}{2} \int \rd x' \rd \tau' \ V\left(x+\frac{x'}{2} \right) V\left(x-\frac{x'}{2} \right)
                           \left\la
                              \tilde{R}^\dagger_{\uparrow} \left( \tau + \frac{\tau'}{2}, x+\frac{x'}{2} \right) \tilde{L}_{\uparrow}\left( \tau + \frac{\tau'}{2}, x+\frac{x'}{2} \right)
                           \right. \times \cr
 && \qquad  \qquad  \qquad  \qquad  \qquad \qquad  \qquad \qquad \times
                            \left.
                              \tilde{R}^\dagger_{\downarrow}\left( \tau - \frac{\tau'}{2}, x-\frac{x'}{2} \right) \tilde{L}_{\downarrow}\left( \tau - \frac{\tau'}{2}, x-\frac{x'}{2} \right)
                            \right\ra  + h.c. \approx \cr
&\approx & \frac{1}{2} V^2(x)  \tilde{R}^\dagger_{\uparrow}(x,\tau) \tilde{L}_{\downarrow}(x,\tau) \times
                             \int \rd x' \rd \tau' \left\la \tilde{R}^\dagger_{\downarrow}\left( \tau - \frac{\tau'}{2}, x-\frac{x'}{2} \right)
                                                                     \tilde{L}_{\uparrow}\left( \tau + \frac{\tau'}{2}, x + \frac{x'}{2} \right) \right\ra  + h.c. \approx \cr
 & \approx & 2 \frac{V(x)^2}{\Delta^{\rm (gen)}} \tilde{R}^\dagger_{\uparrow}(x,\tau) \tilde{L}_{\downarrow}(x,\tau) + h.c.
\eea

%%%
% \input{KL-SU2-SupplMat}
%%%

\end{document}